\documentclass[bibtex,twocolumn,prl,showpacs,floatfix,preprintnumbers,amsmath,amssymb,superscriptaddress,letter]{revtex4}

\usepackage{graphicx}% Include figure files
\usepackage{dcolumn}% Align table columns on decimal point
\usepackage{color}
\usepackage{xspace}
\usepackage[tight]{units}

\newcommand{\numu}{\ensuremath{\nu_{\mu}}\xspace}
\newcommand{\numubar}{\ensuremath{\overline{\nu}_{\mu}}\xspace}

\newcommand{\nutaubar}{\ensuremath{\overline{\nu}_{\tau}}\xspace}

\newcommand{\pt}{\ensuremath{p_{\mathrm{T}}}\xspace}
\newcommand{\pz}{\ensuremath{p_{z}}\xspace}

\newcommand{\piplus}{\ensuremath{\pi^{+}}\xspace}
\newcommand{\piminus}{\ensuremath{\pi^{-}}\xspace}

\newcommand{\muplus}{\ensuremath{\mu^{+}}\xspace}
\newcommand{\muminus}{\ensuremath{\mu^{-}}\xspace}

\newcommand{\eV}{\ensuremath{\mathrm{eV}}\xspace}
\newcommand{\cl}{\ensuremath{\textrm{C.L.}}\xspace}

\newcommand{\mrm}[1]{\mathrm{#1}}
\newcommand{\trm}[1]{\textrm{#1}}

\makeatletter

\newcommand{\Rmnum}[1]{\expandafter\@slowromancap\romannumeral #1@}
\makeatother

\newcommand{\dmbaratm}{\ensuremath{\Delta \overline{m}^{2}}\xspace}
\newcommand{\dmatm}{\ensuremath{\Delta m^{2}}\xspace}
\newcommand{\snthetabar}{\ensuremath{\sin^{2}(2\overline{\theta})}\xspace}
\newcommand{\sntheta}{\ensuremath{\sin^{2}(2\theta)}\xspace}

\newcommand{\systerror}{\ensuremath{^{+10.2}_{-8.9}}\xspace}

\newcommand{\staterrorRHC}{\ensuremath{^{+0.46}_{-0.40}}\xspace}
\newcommand{\selectedEvents}{\ensuremath{130}\xspace}
\newcommand{\predictedEvents}{\ensuremath{136.4\pm11.7\trm{(stat)}\systerror\textrm{(syst)}}\xspace}
\newcommand{\maximalDmConstraint}{\ensuremath{\unit[3.37\times10^{-3}]{\eV^{2}}}\xspace}
\newcommand{\exposure}{\ensuremath{7.1\times 10^{20}}\xspace}

\begin{document}
\addtolength\topmargin{0.5in}

\preprint{FERMILAB-PUB-11-357-PPD}
\preprint{BNL-96122-2011-JA}
\preprint{arXiv:hep-ex/1108.1509}

\vspace*{1.0cm} 
\title{Search for the disappearance of muon antineutrinos in
  the NuMI neutrino beam}
%\title{Author list for May 2011 FHC Nubar oscillation paper (updated 27 July 2011)}

\newcommand{\Berkeley}{Lawrence Berkeley National Laboratory, Berkeley, California, 94720 USA}
\newcommand{\Cambridge}{Cavendish Laboratory, University of Cambridge, Madingley Road, Cambridge CB3 0HE, United Kingdom}
\newcommand{\FNAL}{Fermi National Accelerator Laboratory, Batavia, Illinois 60510, USA}
\newcommand{\RAL}{Rutherford Appleton Laboratory, Science and Technologies Facilities Council, OX11 0QX, United Kingdom}
\newcommand{\UCL}{Department of Physics and Astronomy, University College London, Gower Street, London WC1E 6BT, United Kingdom}
\newcommand{\Caltech}{Lauritsen Laboratory, California Institute of Technology, Pasadena, California 91125, USA}
\newcommand{\Alabama}{Department of Physics and Astronomy, University of Alabama, Tuscaloosa, Alabama 35487, USA}
\newcommand{\ANL}{Argonne National Laboratory, Argonne, Illinois 60439, USA}
\newcommand{\Athens}{Department of Physics, University of Athens, GR-15771 Athens, Greece}
\newcommand{\NTUAthens}{Department of Physics, National Tech. University of Athens, GR-15780 Athens, Greece}
\newcommand{\Benedictine}{Physics Department, Benedictine University, Lisle, Illinois 60532, USA}
\newcommand{\BNL}{Brookhaven National Laboratory, Upton, New York 11973, USA}
\newcommand{\CdF}{APC -- Universit\'{e} Paris 7 Denis Diderot, 10, rue Alice Domon et L\'{e}onie Duquet, F-75205 Paris Cedex 13, France}
\newcommand{\Cleveland}{Cleveland Clinic, Cleveland, Ohio 44195, USA}
\newcommand{\Delhi}{Department of Physics \& Astrophysics, University of Delhi, Delhi 110007, India}
\newcommand{\GEHealth}{GE Healthcare, Florence South Carolina 29501, USA}
\newcommand{\Harvard}{Department of Physics, Harvard University, Cambridge, Massachusetts 02138, USA}
\newcommand{\HolyCross}{Holy Cross College, Notre Dame, Indiana 46556, USA}
\newcommand{\IIT}{Department of Physics, Illinois Institute of Technology, Chicago, Illinois 60616, USA}
\newcommand{\Iowa}{Department of Physics and Astronomy, Iowa State University, Ames, Iowa 50011 USA}
\newcommand{\Indiana}{Indiana University, Bloomington, Indiana 47405, USA}
\newcommand{\ITEP}{High Energy Experimental Physics Department, ITEP, B. Cheremushkinskaya, 25, 117218 Moscow, Russia}
\newcommand{\JMU}{Physics Department, James Madison University, Harrisonburg, Virginia 22807, USA}
\newcommand{\LASL}{Nuclear Nonproliferation Division, Threat Reduction Directorate, Los Alamos National Laboratory, Los Alamos, New Mexico 87545, USA}
\newcommand{\Lebedev}{Nuclear Physics Department, Lebedev Physical Institute, Leninsky Prospect 53, 119991 Moscow, Russia}
\newcommand{\LLL}{Lawrence Livermore National Laboratory, Livermore, California 94550, USA}
\newcommand{\LosAlamos}{Los Alamos National Laboratory, Los Alamos, New Mexico 87545, USA}
\newcommand{\MIT}{Lincoln Laboratory, Massachusetts Institute of Technology, Lexington, Massachusetts 02420, USA}
\newcommand{\Minnesota}{University of Minnesota, Minneapolis, Minnesota 55455, USA}
\newcommand{\Crookston}{Math, Science and Technology Department, University of Minnesota -- Crookston, Crookston, Minnesota 56716, USA}
\newcommand{\Duluth}{Department of Physics, University of Minnesota -- Duluth, Duluth, Minnesota 55812, USA}
\newcommand{\Ohio}{Center for Cosmology and Astro Particle Physics, Ohio State University, Columbus, Ohio 43210 USA}
\newcommand{\Otterbein}{Otterbein College, Westerville, Ohio 43081, USA}
\newcommand{\Oxford}{Subdepartment of Particle Physics, University of Oxford, Oxford OX1 3RH, United Kingdom}
\newcommand{\PennState}{Department of Physics, Pennsylvania State University, State College, Pennsylvania 16802, USA}
\newcommand{\PennU}{Department of Physics and Astronomy, University of Pennsylvania, Philadelphia, Pennsylvania 19104, USA}
\newcommand{\Pittsburgh}{Department of Physics and Astronomy, University of Pittsburgh, Pittsburgh, Pennsylvania 15260, USA}
\newcommand{\IHEP}{Institute for High Energy Physics, Protvino, Moscow Region RU-140284, Russia}
\newcommand{\Rochester}{Department of Physics and Astronomy, University of Rochester, New York 14627 USA}
\newcommand{\RoyalH}{Physics Department, Royal Holloway, University of London, Egham, Surrey, TW20 0EX, United Kingdom}
\newcommand{\Carolina}{Department of Physics and Astronomy, University of South Carolina, Columbia, South Carolina 29208, USA}
\newcommand{\SLAC}{Stanford Linear Accelerator Center, Stanford, California 94309, USA}
\newcommand{\Stanford}{Department of Physics, Stanford University, Stanford, California 94305, USA}
\newcommand{\StJohnFisher}{Physics Department, St. John Fisher College, Rochester, New York 14618 USA}
\newcommand{\Sussex}{Department of Physics and Astronomy, University of Sussex, Falmer, Brighton BN1 9QH, United Kingdom}
\newcommand{\TexasAM}{Physics Department, Texas A\&M University, College Station, Texas 77843, USA}
\newcommand{\Texas}{Department of Physics, University of Texas at Austin, 1 University Station C1600, Austin, Texas 78712, USA}
\newcommand{\TechX}{Tech-X Corporation, Boulder, Colorado 80303, USA}
\newcommand{\Tufts}{Physics Department, Tufts University, Medford, Massachusetts 02155, USA}
\newcommand{\UNICAMP}{Universidade Estadual de Campinas, IFGW-UNICAMP, CP 6165, 13083-970, Campinas, SP, Brazil}
\newcommand{\UFG}{Instituto de F\'{i}sica, Universidade Federal de Goi\'{a}s, CP 131, 74001-970, Goi\^{a}nia, GO, Brazil}
\newcommand{\USP}{Instituto de F\'{i}sica, Universidade de S\~{a}o Paulo,  CP 66318, 05315-970, S\~{a}o Paulo, SP, Brazil}
\newcommand{\Warsaw}{Department of Physics, University of Warsaw, Ho\.{z}a 69, PL-00-681 Warsaw, Poland}
\newcommand{\Washington}{Physics Department, Western Washington University, Bellingham, Washington 98225, USA}
\newcommand{\WandM}{Department of Physics, College of William \& Mary, Williamsburg, Virginia 23187, USA}
\newcommand{\Wisconsin}{Physics Department, University of Wisconsin, Madison, Wisconsin 53706, USA}
\newcommand{\deceased}{Deceased.}

\affiliation{\ANL}
\affiliation{\Athens}
%\affiliation{\Benedictine}
\affiliation{\BNL}
\affiliation{\Caltech}
\affiliation{\Cambridge}
\affiliation{\UNICAMP}
%\affiliation{\CdF}
\affiliation{\FNAL}
\affiliation{\UFG}
\affiliation{\Harvard}
\affiliation{\HolyCross}
\affiliation{\IIT}
\affiliation{\Indiana}
\affiliation{\Iowa}
%\affiliation{\IHEP}
%\affiliation{\ITEP}
%\affiliation{\JMU}
%\affiliation{\Lebedev}
%\affiliation{\LLL}
\affiliation{\UCL}
\affiliation{\Minnesota}
\affiliation{\Duluth}
\affiliation{\Otterbein}
\affiliation{\Oxford}
\affiliation{\Pittsburgh}
\affiliation{\RAL}
\affiliation{\USP}
\affiliation{\Carolina}
\affiliation{\Stanford}
\affiliation{\Sussex}
\affiliation{\TexasAM}
\affiliation{\Texas}
\affiliation{\Tufts}
\affiliation{\Warsaw}
%\affiliation{\Washington}
\affiliation{\WandM}
%\affiliation{\Wisconsin}

\author{P.~Adamson}
\affiliation{\FNAL}
%\affiliation{\UCL}
%\affiliation{\Sussex}

%\author{C.~Andreopoulos}
%\affiliation{\RAL}
%\affiliation{\Athens}

%\author{K.~E.~Arms}
%\affiliation{\Minnesota}

%\author{R.~Armstrong}
%\affiliation{\Indiana}

\author{D.~J.~Auty}
\affiliation{\Sussex}

%\author{S.~Avvakumov}
%\affiliation{\Stanford}

\author{D.~S.~Ayres}
\affiliation{\ANL}

\author{C.~Backhouse}
\affiliation{\Oxford}

%\author{B.~Baller}
%\affiliation{\FNAL}

%\author{B.~Barish}
%\affiliation{\Caltech}

%\author{P.~D.~Barnes~Jr.}
%\affiliation{\LLL}

\author{G.~Barr}
\affiliation{\Oxford}

%\author{W.~L.~Barrett}
%\affiliation{\Washington}

%\author{E.~Beall}
%\altaffiliation[Now at\ ]{\Cleveland .}
%\affiliation{\ANL}
%\affiliation{\Minnesota}

%\author{B.~R.~Becker}
%\affiliation{\Minnesota}

%\author{A.~Belias}
%\affiliation{\RAL}

%\author{R.~H.~Bernstein}
%\affiliation{\FNAL}

%\author{M.~Betancourt}
%\affiliation{\Minnesota}

%\author{D.~Bhattacharya}
%\affiliation{\Pittsburgh}

%\author{M.~Bhattarai}
%\affiliation{\Texas}
%\affiliation{\Duluth}

\author{M.~Bishai}
\affiliation{\BNL}

\author{A.~Blake}
\affiliation{\Cambridge}

%\author{B.~Bock}
%\affiliation{\Duluth}

\author{G.~J.~Bock}
\affiliation{\FNAL}

\author{D.~J.~Boehnlein}
\affiliation{\FNAL}

\author{D.~Bogert}
\affiliation{\FNAL}

%\author{P.~M.~Border}
%\affiliation{\Minnesota}

%\author{C.~Bower}
%\affiliation{\Indiana}

%\author{E.~Buckley-Geer}
%\affiliation{\FNAL}

% ...

\collaboration{The MINOS Collaboration}
\noaffiliation

%\date{\today}

%\maketitle

%\end{document}

\begin{abstract}

We report constraints on antineutrino oscillation parameters that were
obtained by using the two MINOS detectors to measure the $7\%$ muon
antineutrino component of the NuMI neutrino beam. In the Far Detector,
we select \selectedEvents events in the charged-current muon
antineutrino sample, compared to a prediction of \predictedEvents
events under the assumption $|\dmbaratm|=2.32\times10^{-3}~\eV^{2}$,
$\snthetabar=1.0$. Assuming no oscillations occur at the Near Detector
baseline, a fit to the two-flavor oscillation approximation constrains
$|\dmbaratm|<\maximalDmConstraint$ at the 90\% confidence level with
$\snthetabar=1.0$.

\end{abstract}
 \maketitle

%*********************************************************************72
%			Introduction Section
%*********************************************************************72

%PRD RC guidance
%http://prd.aps.org/info/polprocd.html
%Rapid Communications are intended for important new results which
%deserve accelerated publication, and are therefore given priority in
%editorial processing and production to minimize the time between
%receipt and publication. Rapid Communications are similar to Physical
%Review Letters; the principal difference is that Letters are
%accessible to a general audience of physicists and allied scientists,
%while Rapid Communications are primarily for a more specialized
%audience, the usual readers of Physical Review D. Rapid Communications
%in Physical Review D are limited to five journal pages.

The phenomenon of neutrino oscillations has been well established by
experimental
observations~\cite{ref:minos2006,ref:sk,ref:borexino,ref:soudan2,ref:macro,ref:k2k,ref:sno,ref:kamland}. The
underlying quantum-mechanical mixing between the neutrino flavor and
mass eigenstates is governed by the elements of the PMNS
matrix~\cite{ref:pmns}, usually parameterized by three mixing angles
and a CP-violating phase. Oscillations are governed by the ratio of
the distance traveled by the neutrino to its energy ($L/E$) and the
two independent neutrino mass-squared differences. CPT symmetry
constrains the allowed differences between a particle and its
antiparticle~\cite{ref:weinberg} and requires their masses to be
identical. Differences between the measured neutrino and antineutrino
oscillation parameters would indicate new physics. For example, as
neutrinos propagate through matter, nonstandard
interactions~\cite{ref:NSI} could alter the disappearance
probabilities of neutrinos relative to antineutrinos and thus the
inferred oscillation parameters~\cite{ref:minosNSIfits}. Such models
of new physics predict a different energy dependence and so probing
the standard oscillation hypothesis to greater precision across a wide
range of energies is valuable.

The MINOS long-baseline experiment has made the most precise
measurements to date of the larger (atmospheric) mass-squared
splitting for both neutrinos~\cite{ref:minosCC2010} and
antineutrinos~\cite{ref:rhc}. With the NuMI facility~\cite{ref:NuMI}
configured to provide a neutrino-dominated beam, a measurement of
\numu disappearance resulted in a mass-squared splitting of
$|\dmatm|=(2.32^{+0.12}_{-0.08})\times10^{-3}$~eV$^2$ and mixing angle
$\sntheta>0.90$ (90\% confidence limit
      [C.L.])~\mbox{\cite{ref:minosCC2010,comment:dmbar}}. From direct
      observations of \numubar disappearance, using a smaller exposure
      to the beam optimized for antineutrinos, MINOS measures the
      antineutrino oscillation parameters
      $|\dmbaratm|=[3.36\staterrorRHC
        \trm{(stat)}\pm0.06\trm{(syst)}]\times10^{-3}\ \eV^{2}$ and
      $\snthetabar=0.86^{+0.11}_{-0.12}\trm{(stat)}\pm0.01\trm{(syst)}$~\cite{ref:rhc}. Prior
      to the measurement of $|\dmbaratm|$ by MINOS the strongest
      constraints on antineutrino oscillation parameters came from a
      fit~\cite{ref:maltoni} to global data dominated by
      Super-Kamiokande results where the sum of atmospheric \numu and
      \numubar interaction rates was measured.

This paper describes an analysis of the 7\% $\numubar$ component of
the NuMI beam, optimized to produce neutrinos, with an exposure of
\exposure protons on target. The MINOS detectors are magnetized,
allowing event-by-event separation of \numu and \numubar
charged-current (CC) events using the curvature of the muon track.
The \numubar sample presented here provides a new test of the
oscillation hypothesis for muon antineutrinos at the atmospheric
scale. With substantially increased statistics in the 5--15~GeV energy
range relative to the sample obtained with the beam configured for
antineutrinos~\cite{ref:rhc} the \numubar oscillation probability can
be probed to greater precision in this region.

The NuMI beam uses 120~GeV/c protons incident on a graphite target to
produce secondary hadrons, in particular pions and kaons of both
charges. Depending on the sign of the applied current, two magnetic
horns focus either positively or negatively charged hadrons for a
neutrino or antineutrino beam, respectively. A 675~m long iron-walled
decay pipe --- evacuated during the first half of the data taking
period but later filled with 0.9~atm helium for structural reasons ---
allows the hadrons and tertiary muons to decay in flight, producing
neutrinos and antineutrinos. The antineutrino component of the
neutrino beam arises from four main sources: decays of hadrons
traveling along the axes of the horns where the focusing field is
negligible; partially defocused hadrons decaying close to the horns;
decays of hadrons produced from interactions with the helium and walls
of the decay pipe; and decays of tertiary muons that arise mainly from
decays of the focused hadrons. Muon antineutrinos from neutral kaons
are estimated from simulation to comprise 0.6\% of events across the
spectrum. The combined energy spectrum of the \numubar CC events
arising from these sources is broadly distributed and peaks at
approximately 8~GeV, whereas the energy spectrum resulting from the
focused hadrons is narrowly-peaked at approximately 3~GeV.

The two MINOS detectors~\cite{ref:NIM} are located
\mbox{\unit[1.04]{km}} [Near Detector (ND)] and \mbox{\unit[735]{km}}
     [Far Detector (FD)] from the target. Both detectors are segmented
     steel/scintillator tracking calorimeters. The detector fiducial
     masses are 23.7~tons and 4.2~kilotons at the ND and FD
     respectively. In CC interactions,
     $\numu(\numubar)+\mrm{N}\rightarrow\mu^-(\mu^+)+X$, a hadronic
     shower~($X$) and a muon track may be observed. The reconstructed
     neutrino energy is the sum of the reconstructed muon and hadron
     energies. Hadronic energy is measured by calorimetry. Muon energy
     is measured by range for contained tracks or by curvature in a
     \unit[1.4]{T} toroidal magnetic field for exiting tracks. For
     this data set, the fields in both detectors have been set so that
     they focus \muminus and defocus \muplus, allowing the separation
     of \numu and \numubar~CC interactions.

The inclusive \numubar CC interaction rate as a function of
reconstructed \numubar energy is measured in each detector. The
measured FD spectrum is compared to the projection of the ND data to
the FD, taking into account the different geometric acceptances of the
two detectors. In this comparison, many sources of systematic
uncertainty largely cancel due to the similarities of the two
detectors. Antineutrino oscillations would cause an energy-dependent
\numubar deficit at the FD compared to the projection from the ND; the
\numubar survival probability in the two-flavor approximation is
\begin{equation}
P(\numubar\rightarrow\numubar) =
1 - \snthetabar
\sin^{2}\left(
\frac{1.267\dmbaratm L}{E}
\right),
\label{prob}
\end{equation}
where $L$~[km] is the distance from the point of antineutrino
production, $E$~[GeV] the antineutrino energy, $\overline\theta$ the
antineutrino mixing angle, and $\Delta\overline{m}^2$~[$\eV^{2}$] the
antineutrino mass-squared difference.

\begin{figure}
\centering
\includegraphics[trim = 10mm 25mm 10mm 10mm, clip, width=\columnwidth]{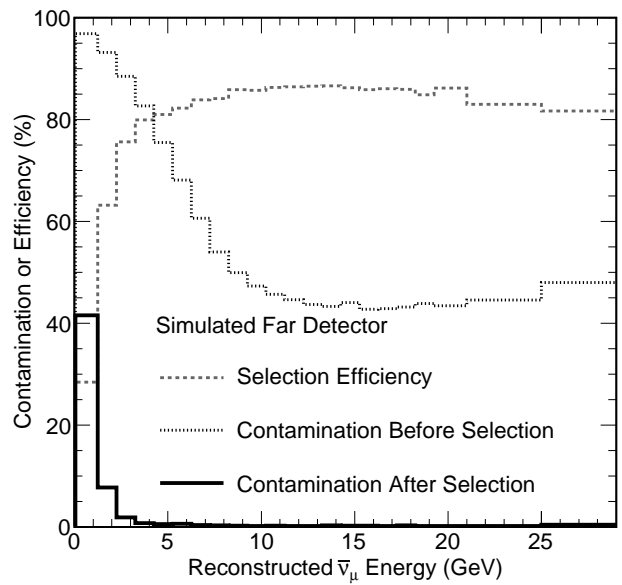}
\caption{Efficiency of the selection of $\bar{\nu}_\mu$ CC candidate
  events reconstructed with a positive charge-sign track in the Far
  Detector. The contamination due to misidentified NC and $\nu_\mu$~CC
  interactions is also shown (assuming no oscillations), both before
  and after all other selection criteria are applied.}
\label{fig:effcont} 
\end{figure}

Selected events must contain at least one reconstructed track; the
longest track is identified as the muon candidate. This muon candidate
must originate inside the fiducial volume and have a positive charge
determined from track curvature. However, the track finding algorithm
can occasionally form a track out of hadronic activity, or misidentify
the curvature of a muon track. A simple charge-sign selection based on
this track-fit information yields a sample that is highly contaminated
with both \numu CC and neutral current (NC) events as shown in
\mbox{Fig. \ref{fig:effcont}}. Monte Carlo studies show that about
half of NC events with a reconstructed track and 7\% of \numu CC
events with a track are misidentified as $\mu^{+}$ candidates. Most of
the misidentified \numu CC events are high-inelasticity interactions
in which the soft $\mu^{-}$ is obscured by the hadronic shower.  In
addition, higher momentum muons follow a less curved trajectory,
increasing the probability of charge misidentification. With the beam
consisting of about 92\% muon neutrinos, the initial signal to
background ratio is inherently much lower for muon antineutrinos than
it is for neutrinos and the development of further selection cuts was
necessary.

To reduce the misidentified NC and $\numu$ CC background events, three
selection variables are used. The first is a likelihood-based
separation parameter based on event topology. The second variable is a
measure of the confidence of charge-sign determination from the track
fitting. The third variable provides an additional measure
of the direction of curvature of the muon track by comparing the local
track direction at the vertex to that at the end point of the
track~\cite{ref:RustemThesis}. The likelihood-based separation
parameter was originally developed to distinguish NC background from
\numu CC events in the MINOS analysis of $\nu_\mu$
oscillations~\cite{ref:minos2006} but it is also effective in removing
the misidentified high-inelasticity $\numu$ CC background. This
discriminator uses probability density functions constructed from
three variables: the event length, the fraction of the total event
signal in the reconstructed track, and the average signal per plane of
the reconstructed track. These quantities are related to the muon
range, the event inelasticity and the average energy loss
$\mathrm{d}E/\mathrm{d}x$ of the muon track and are distributed
differently for $\numubar$~CC events compared to NC and misidentified
$\numu$~CC events.

The selection was optimized~\cite{ref:DavidThesis} for statistical
sensitivity to oscillation parameters equal to those measured for
$\numu$~\cite{ref:minos2008}. Figure~\ref{fig:effcont} shows the
efficiency of the full selection and the remaining contamination as a
function of \numubar energy in the FD\@. Assuming no oscillations, the
efficiency of the selection is 85\% and the purity of the \numubar CC
sample is 98\%, integrated over all energies in the FD\@.

\begin{figure}
\centering
\includegraphics[trim = 10mm 25mm 10mm 10mm, clip, width=\columnwidth]{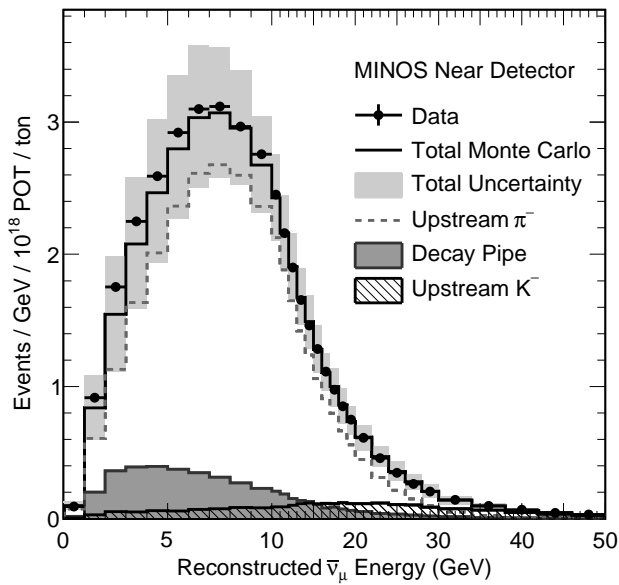}
\caption{Reconstructed energy spectra of \numubar CC candidate events
  at the Near Detector. The solid line shows the Monte Carlo
  simulation, which is broken into three sources of \numubar parent
  particles. The upstream pion decay contribution originates primarily
  from the target but also includes antineutrinos from muons whose
  parents decayed in the upstream region. The decay pipe component
  corresponds to all \numubar parents (other than muons, which
  contribute about 3\% of the ND spectrum) produced 45~m or more
  downstream of the target. The contribution from kaon decay is shown
  by the hatched histogram. The shaded band on the simulation shows
  the size of the systematic error on the absolute ND spectrum
  prediction.}
\label{fig:flux}
\end{figure}

The measured ND energy spectrum, shown in Fig.~\ref{fig:flux}, is used
to predict the FD spectrum, as in previous MINOS
analyses~\cite{ref:minos2006, ref:minos2008, ref:minosCC2010,
  ref:JustinThesis}. This effectively mitigates sources of
mismodeling, such as uncertainties in the neutrino flux or neutrino
cross sections, which affect both detectors in similar ways.

Hadron production in the NuMI target and beam line is simulated with
\mbox{\sc FLUKA \cite{ref:fluka}} by using {\sc
  FLUGG}~\cite{ref:flugg} as an interface to the {\sc
  GEANT4}~\cite{ref:G4} based geometry. Additionally, hadron
production in the target is constrained by a fit to ND
\mbox{spectra\,\cite{ref:minos2006}}, which correct the $\pi$ and $K$
distributions as a function of their transverse and longitudinal
momenta at production, \pt and \pz respectively. The fit is performed
simultaneously for several different beam configurations, which
permits the constraint of a wide range of \pt-\pz space for \numu
parent particles. The $\piplus/\piminus$ ratio measured by
NA49~\cite{ref:NA49}, together with the \pt spectral shape from the
\numu fit, constrains the \numubar parent \pt spectral shape, while a
fit to the ND \numubar energy spectrum provides overall normalization
and \pz shape information.  These fit parameters have been applied to
the flux in obtaining the simulated ND spectrum shown in
\mbox{Fig.\ \ref{fig:flux}}. The errors obtained in the fit provide an
estimate of the uncertainty on the hadron production from the target;
the corresponding error on the FD event rate, extrapolated from ND
data, is less than 1\% for the beam component that arises directly
from hadrons produced in the target.

Figure~\ref{fig:flux} shows the contribution of different beam flux
components to the \numubar CC interaction rate in the ND as a function
of energy. A significant fraction of ND events originate from parent
particles produced in the decay pipe, predominantly from interactions
of primary and secondary hadrons with the decay pipe walls and the
helium (muons are not included in our decay pipe component definition
as they are constrained by the ND $\numu$~CC events). For these events
the relative acceptance of the ND compared to the FD is larger than
for particles produced in upstream interactions. Consequently, the
contribution from decay pipe parent particles as a fraction of the
total spectrum is larger at the ND (12\%) compared to the FD (7\%,
assuming no oscillations). A systematic uncertainty on the size of the
decay pipe component was assessed by scaling this component in the
Monte Carlo simulation and comparing with the ND data. Conservative
scale factors of $\pm$100\% are applied to the decay pipe component,
introducing an uncertainty on the total \numubar CC interaction rate
predicted at the FD of~$^{+6.2}_{-5.0}\%$.

Further systematic uncertainties include a 4\% relative normalization
uncertainty between the ND and FD to account for uncertainties in the
reconstruction efficiencies, exposure and fiducial masses of both
detectors~\cite{ref:minos2008}. A comparison of momentum measurement
from curvature vs.\ range in stopping muon tracks constrains the
uncertainty in track momentum determination from curvature to be
$4\%$. The 50\% uncertainty on the misidentified NC and $\nu_{\mu}$~CC
events was estimated by scaling those components in the ND until the
MC matched the data for the set of events that narrowly failed the
selection on the likelihood-based separation parameter. The total
systematic uncertainty on the predicted number of events at the FD is
82\% of the total statistical uncertainty, assuming oscillation
parameters equal to those measured for \numu~\cite{ref:minosCC2010}.

\begin{figure}
\centering
\includegraphics[trim = 10mm 25mm 10mm 10mm, clip, width=\columnwidth]{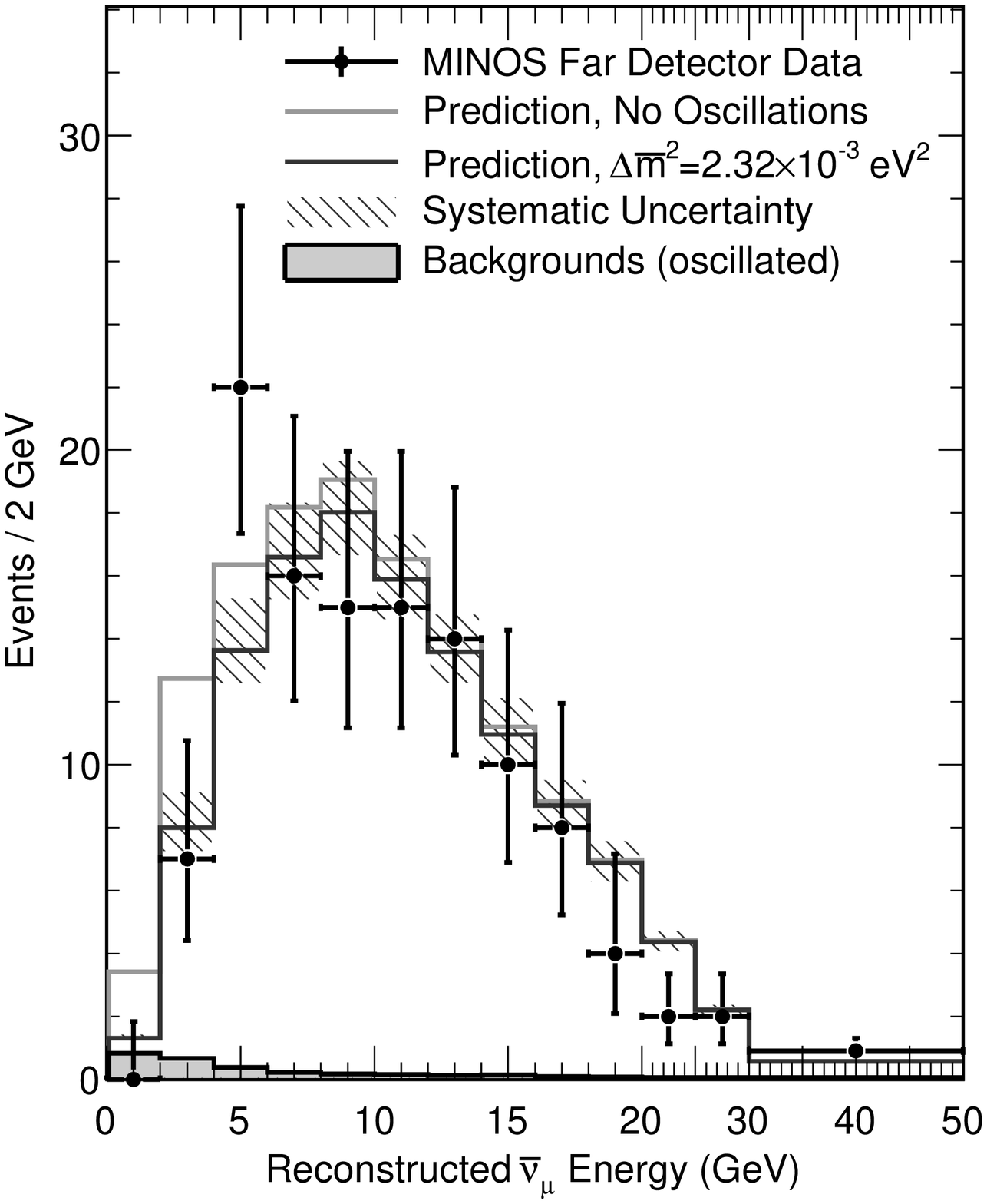}
\caption{Energy spectra of \numubar CC candidate events observed in
  the Far Detector. The predicted spectrum with no oscillations and
  with oscillation parameter values of
  $|\dmbaratm|=|\dmatm|=2.32\times10^{-3}\,\eV^{2}$,
  $\snthetabar=\sntheta=1.0$ are overlaid. The hatched band indicates
  the total systematic uncertainty on the prediction. The estimated
  background includes oscillations at the best-fit values determined
  by the MINOS \numu CC disappearance analysis~\cite{ref:minosCC2010}
  for the \numu CC events.}
\label{fig:energy} 
\end{figure}

At the FD a total of \selectedEvents selected \numubar CC candidate
events are observed. Figure~\ref{fig:energy} shows the energy spectrum
of the FD data overlaid with two predicted spectra obtained from the
ND data: one without oscillations and one with oscillation parameters
of $|\dmbaratm|=2.32\times10^{-3}\ \eV^{2}$,
$\snthetabar=1.0$~\cite{ref:minosCC2010}. The predicted backgrounds
are 1.8 \numu~CC events, 1.2 NC events and 0.2 \nutaubar~CC events (in
the oscillated case). The integrated number of events observed and
expected are detailed in Table~\ref{tab:runs}. The number of FD events
measured in run periods \Rmnum{1} and \Rmnum{2} is smaller than the
prediction. In run period \Rmnum{3}, which differs due to the helium
in the decay pipe, a larger number of events are measured compared
with the prediction. The probability of observing a comparable or
larger difference in event rate between the two periods, evaluated
using mock Monte Carlo experiments, is 8.4\%.

\begin{table}
\begin{center}
\centering
\begin{tabular}{ | c | c | | c | c | c | }
\hline 
Run &  POT &  Events &  Events &  Events \rule{0pt}{2.7ex} \\ 
period &  ($10^{20}$) &  \,observed\, &  expected &  ~expected~ \\
 &   &  &  (oscillated) &  (no osc.) \\[2pt] \hline

\Rmnum{1} \& \Rmnum{2} & 3.21 & 43 & $60.2^{+8.7}_{-8.5}$ & $66.4^{+9.2}_{-9.0}$ \rule{0pt}{3.0ex} \\[4pt]

\Rmnum{3} & 3.88 & 87 & $76.2^{+10.9}_{-10.2}$ & $83.9^{+11.6}_{-10.9}$ \rule{0pt}{3.0ex} \\[4pt] \hline

Total & 7.09 & \selectedEvents & $136.4^{+15.5}_{-14.7}$ & $150.3^{+16.6}_{-15.6}$ \rule{0pt}{3.0ex} \\[2pt] \hline
\end{tabular}
\hfill \hfill \end{center}
\caption{Candidate \numubar CC events observed and expected in the Far
  Detector, broken down into two periods of approximately equal
  exposure. The expected number of events in the oscillated case uses
  the parameters measured with the \numu CC
  sample~\cite{ref:minosCC2010}.}
\label{tab:runs}
\end{table}

\begin{figure}
\centering \includegraphics[trim = 10mm 35mm 10mm 10mm, clip,
  width=\columnwidth]{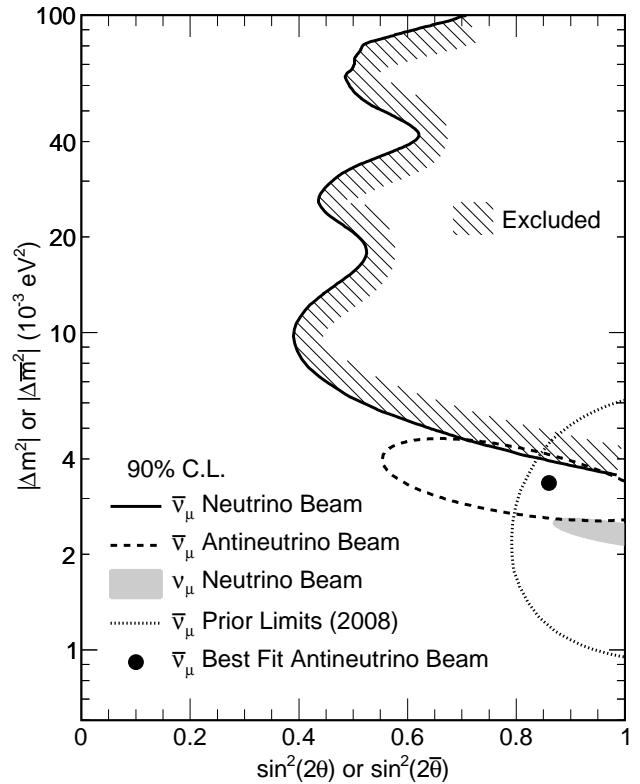}
\caption{Allowed regions for \numubar oscillation parameters from a
  fit to the data in Fig.~\ref{fig:energy}. The region indicated by
  the hashing is excluded. Shown alongside are contours for: the MINOS
  $\numubar$ result from the NuMI beam optimized for antineutrino
  production~\cite{ref:rhc}; the MINOS allowed region for
  neutrinos~\cite{ref:minosCC2010}; and limits from a
  fit~\cite{ref:maltoni} prior to all MINOS \numubar data.}
\label{fig:limitplot}
\end{figure}

The measured FD energy spectrum is compared to that predicted from the
ND assuming $\numubar\rightarrow\nutaubar$ oscillations, following
Eq.~(\ref{prob}). This comparison is made by minimizing a binned
log-likelihood with respect to \dmbaratm and \snthetabar. The
Feldman-Cousins approach~\cite{ref:FeldCous} is used to obtain
confidence limits on the oscillation parameters with systematic
uncertainties included~\cite{ref:NickThesis,ref:AlexThesis}. The
confidence limits thus obtained are shown in
Fig.~\ref{fig:limitplot}. Values of $|\dmbaratm|$ greater than
1~eV$^2$ are not considered in this analysis, since above that point
oscillations with maximal mixing would cause more than 1\% of the
$\numubar$ to disappear in the ND\@. Figure~\ref{fig:limitplot} also
shows the recent MINOS result using the beam configured for
antineutrinos~\cite{ref:rhc}, the MINOS allowed region for
neutrinos~\cite{ref:minosCC2010}, and a fit~\cite{ref:maltoni} to all
global data available prior to all MINOS \numubar data. The MINOS data
presented in this paper are consistent with both the previous MINOS
neutrino and antineutrino limits, and with the limits from a global
fit~\cite{ref:maltoni}. A $\chi^2$ goodness-of-fit test using the
oscillation parameters from~\cite{ref:minosCC2010} yields a
probability of 18\%. Under the assumption $\snthetabar=1.0$ these data
constrain $|\dmbaratm|<\maximalDmConstraint$ (90\%~\cl) in the
two-flavor approximation.

In summary, a high-purity sample of muon antineutrino charged-current
events was selected in the MINOS data from the 7\% \numubar component
of the NuMI neutrino beam. At the Far Detector, \selectedEvents
\numubar event candidates were observed, which is consistent with the
predicted rate in the case of oscillations of \predictedEvents under
the assumption $|\dmbaratm|=2.32\times10^{-3}~\eV^{2}$,
$\snthetabar=1.0$. These data provide a new probe of the oscillation
hypothesis for muon antineutrinos at the atmospheric
scale. Significantly increased statistics in the 5--15~GeV energy
range, compared to the \numubar sample obtained with the NuMI beam
configured for antineutrinos, have allowed the oscillation probability
to be measured with greater precision in this region and have added to
constraints on antineutrino oscillation parameters.

This work was supported by the U.S. DOE; the United Kingdom STFC; the
U.S. NSF; the State and University of Minnesota; the University of
Athens, Greece; and Brazil's FAPESP and CNPq.  We are grateful to the
Minnesota Department of Natural Resources, the crew of the Soudan
Underground Laboratory, and the personnel of Fermilab for their
contribution to this effort.

%*********************************************************************72
%*********************************************************************72
%*********************************************************************72


\begin{thebibliography}{99}
\bibitem{comment:dmbar} The experiment measures an unresolved mixture
  of $|\Delta m^{2}_{31}|$ and $|\Delta m^{2}_{32}|$, which is
  referred to as $|\Delta m^{2}|$ for brevity. The parameter
  $\sin^{2}(2\theta)$ is likewise an admixture, dominated by
  $\theta_{23}$. Similarly for $|\Delta\overline{m}^{2}|$ and
  $\sin^{2}(2\overline{\theta})$.
\end{thebibliography}
\end{document}